\documentclass{ismdproc}

\begin{document}
\title{Understanding RHIC Collisions: Modified QCD fragmentation vs quark coalescence from a thermalized flowing medium}

\def\bea{\begin{eqnarray}}
\def\eea{\end{eqnarray}}

\def\pp{\mbox{$p$-$p$} }
\def\auau{\mbox{Au-Au} }
\def\cucu{\mbox{Cu-Cu} }
\def\pbpb{\mbox{Pb-Pb} }
\def\aa{\mbox{A-A} }
\def\nn{\mbox{N-N} }
\def\ss{\mbox{S-S} }
\def\ee{\mbox{$e^+$-$e^-$} }
\def\ppbar{\mbox{$p$-$\bar p$} }
\def\qqbar{\mbox{$q$-$\bar q$} }
\def\pt{$p_t$ }


%

%

%
\author{{\slshape Tom Trainor}\\[1ex]
CENPA 354290, University of Washington, Seattle, Washington 98195, USA }

\contribID{xy}  
\confID{yz}
\acronym{ISMD2010}
\doi            

\maketitle

\begin{abstract}
The hydrodynamic (hydro) model applied to data from the relativistic heavy ion collider (RHIC) suggests that a dense QCD medium opaque to partons is formed in more-central \auau collisions.  
However, two-component spectrum analysis reveals a hard component, consistent with parton fragmentation described by pQCD which can masquerade as ``radial flow.'' 
Minimum-bias angular correlations reveal that most scattered partons survive as ``minijets'' even in central \auau collisions. 
Such alternative methods quantitatively describe spectrum and correlation structure via pQCD calculations.  RHIC collisions appear to be dominated by parton scattering and fragmentation even in central \auau collisions. 
\end{abstract}

\section{Introduction}

We wish to test the extent to which perturbative QCD (pQCD) can describe more-central \aa collisions at RHIC. Is a hydrodynamic (hydro) description necessary, or even allowed by data? 
Detailed arguments are presented in Refs.~\cite{hardspec,fragevo,tzyam,nohydro} with related material on hydro interpretations in Refs.~\cite{quadspec,davidhq,davidhq2,davidaustin}.
We describe pQCD fragment distributions (FDs) obtained by folding measured fragmentation functions (FFs) with a pQCD dihadron spectrum.
 We adopt a parton ``energy-loss'' model~\cite{bw} to provide FD calculations which describe fragmentation evolution with \aa collision centrality. 
We also introduce a method to convert jet angular correlations to fragment yields and spectra. It is then possible to calculate the minijet contribution to the \aa final state and provide a comprehensive pQCD description of RHIC collisions.

Measurement of azimuth quadrupole $v_2$ via nongraphical numerical methods has led to strong conclusions about hydrodynamic evolution of a thermalized bulk medium with small viscosity. However, trends obtained from an alternative 2D $v_2$ method suggest that the azimuth quadrupole is a novel QCD phenomenon carried by a small fraction of all final-state particles.

\section{Spectra and parton fragmentation}

The two-component model for per-participant-pair \aa spectra~\cite{hardspec,ppprd} is
\bea  \label{aa2comp}
\frac{2}{n_{part}} \frac{1}{y_t}\frac{dn_{ch}}{dy_t} &=& S_{NN}(y_t) +  \nu\, H_{AA}(y_t,\nu) \\ \nonumber
&=&  S_{NN}(y_t) +  \nu\,r_{AA}(y_t,\nu) \,H_{NN}(y_t),
\eea
where $S_{NN}$ is the soft component and $H_{AA}$ is the \aa hard component (with reference $H_{NN}\sim H_{pp}$). Ratio $r_{AA} = H_{AA} / H_{NN}$ is an alternative to nuclear modification factor $R_{AA}$. Centrality measure $\nu \equiv 2 n_{binary} / n_{participant}$ estimates the mean projectile-nucleon path length in \aa collisions. 
Soft component $S_{NN}$ by hypothesis remains unchanged with \aa centrality. 
For Glauber linear superposition of \pp (N-N) collisions (the GLS reference) spectrum hard component $H_{AA} \rightarrow H_{NN}(y_t)$ also remains unchanged. In more-central \aa collisions $H_{AA}(y_t,b)$ deviates from reference $H_{NN}(y_t)$, reflecting ``medium modification'' of parton fragmentation.
%
\begin{figure*}[t]
\includegraphics[width=0.24\textwidth,height=.25\textwidth]{
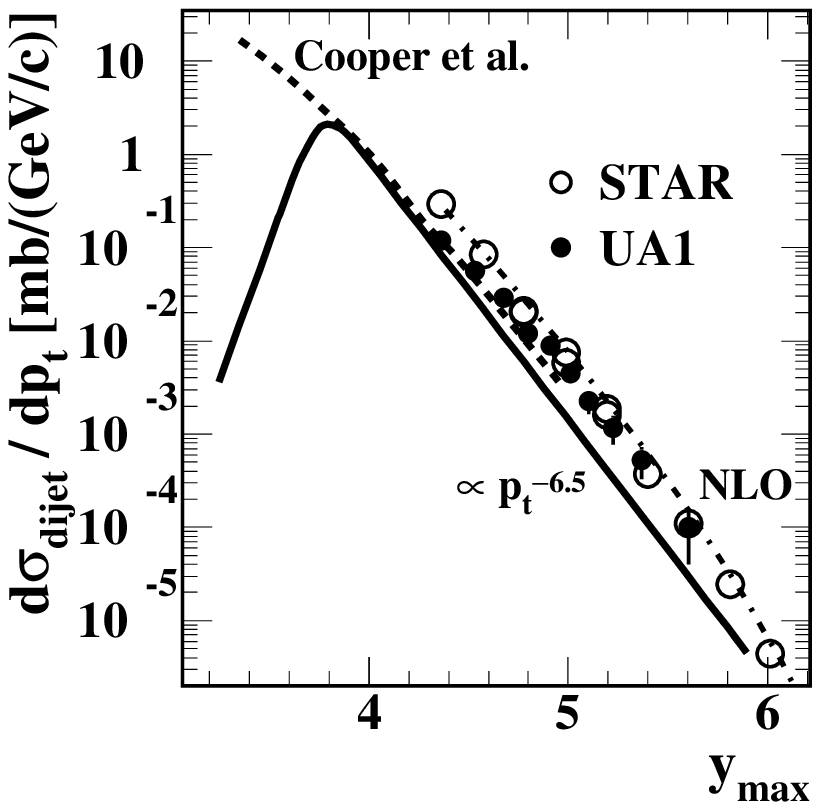} 
\includegraphics[width=0.24\textwidth,height=.255\textwidth]{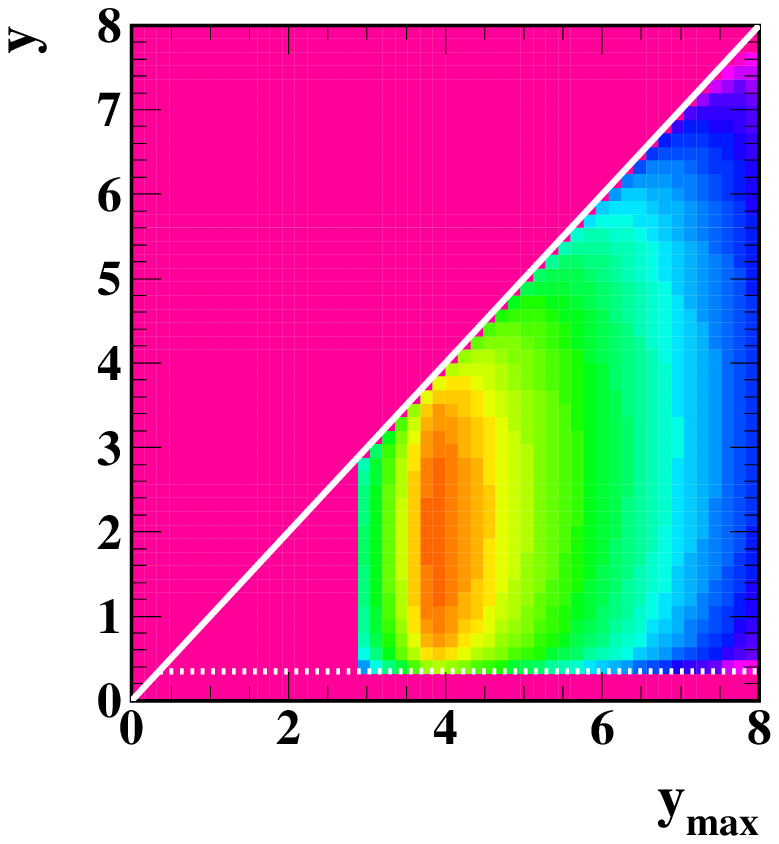}
\includegraphics[width=0.24\textwidth,height=.25\textwidth]{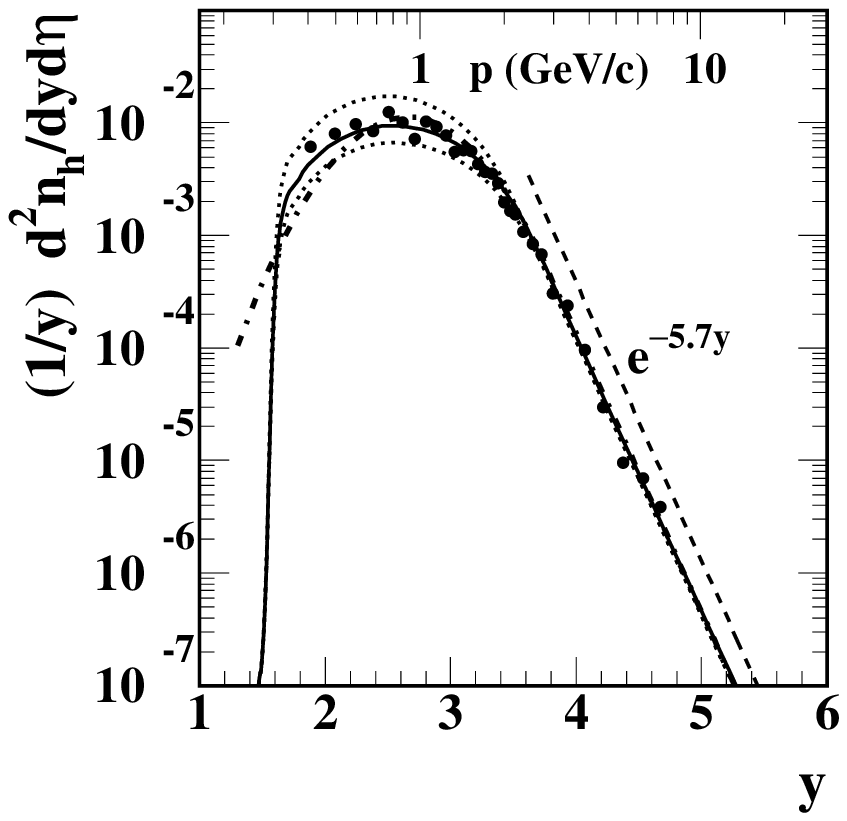}
\includegraphics[width=.25\textwidth,height=0.25\textwidth]{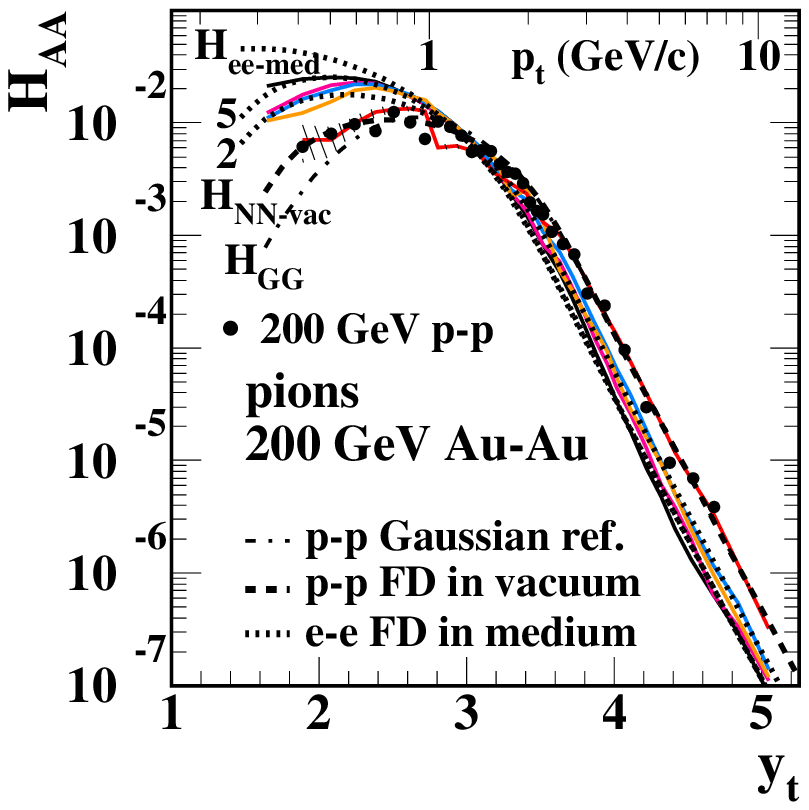}
\caption{\label{trainor_tom.fig1}
First: Dijet spectrum for \pp collisions (solid curve) compared to an ab-initio pQCD theory result (bold dotted curve) and reconstructed jets (points)~\cite{fragevo},
Second: Eq.~(\ref{fold}) integrand for $e^+$-$e^-$ FFs,
Third: Fragment distribution (solid curve) compared to \pp hard-component data (points)~\cite{ppprd}. 
Fourth: Medium-modified FDs compared to \auau hard components~\cite{fragevo}. 
}
\end{figure*}

We adopt the hypothesis that spectrum hard component $H_{AA}$ represents minimum-bias parton scattering and fragmentation in the form of a {\em fragment distribution} (FD). The pQCD convolution integral used to calculate fragment distributions is
\bea \label{fold}
\frac{d^2n_{h}}{dy_t\, d\eta} = 2\pi y_t H_{AA}(y_t)   &\approx&    \frac{\epsilon(\Delta \eta)/2}{ \sigma_{_{\tiny NSD}}\, \Delta \eta_{4\pi}}  \int_0^\infty   dy_{max}\, D_{xx}(y_t,y_{max}) \frac{d\sigma_{dijet}}{dy_{max}},
\eea
where $D_{xx}(y_t,y_{max})$ is the measured FF ensemble  for collision system $xx$ and $d\sigma_{dijet}/dy_{max}$ is the pQCD dijet (parton) spectrum~\cite{fragevo}.  The folding integral then predicts hadron spectrum hard component ${d^2n_{h}}/{dy_t\, d\eta}$ from parton pairs  scattered into  angular acceptance $\Delta \eta$. Comparisons with measured spectrum hard components in \pp and \auau collisions are shown in Fig.~\ref{trainor_tom.fig1}. The two-component model appears to provide a full description of \aa spectra in terms of pQCD.

\section{Correlations and parton fragmentation}

 \begin{figure*}[h]
\includegraphics[width=.24\textwidth,height=0.25\textwidth]{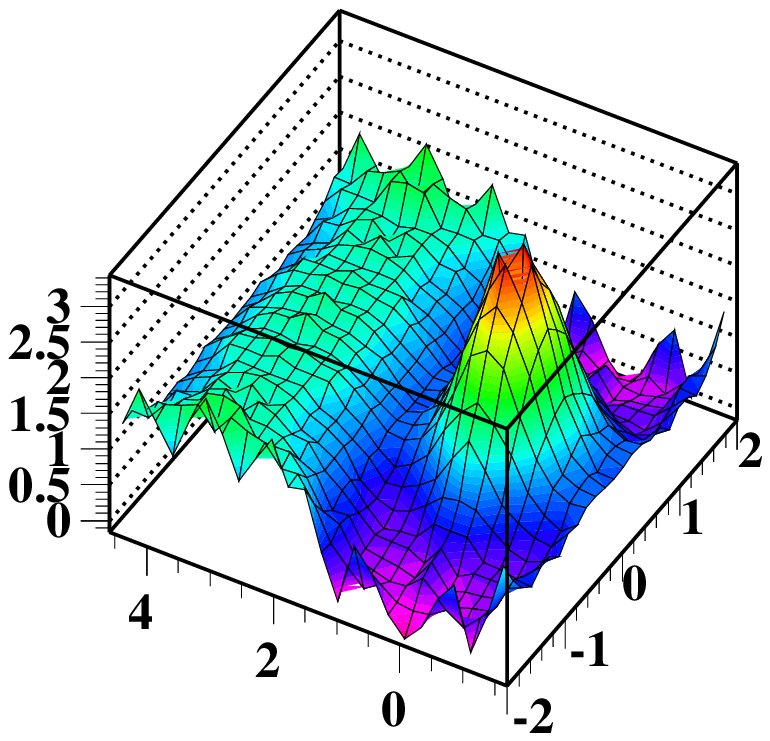}
  \includegraphics[width=.24\textwidth,height=0.25\textwidth]{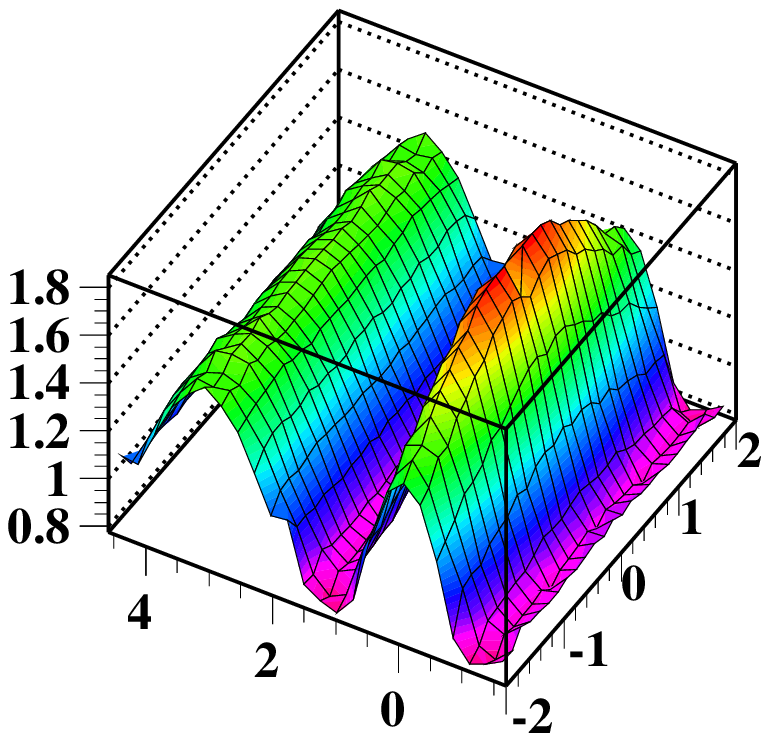}
 \includegraphics[width=.24\textwidth,height=0.25\textwidth]{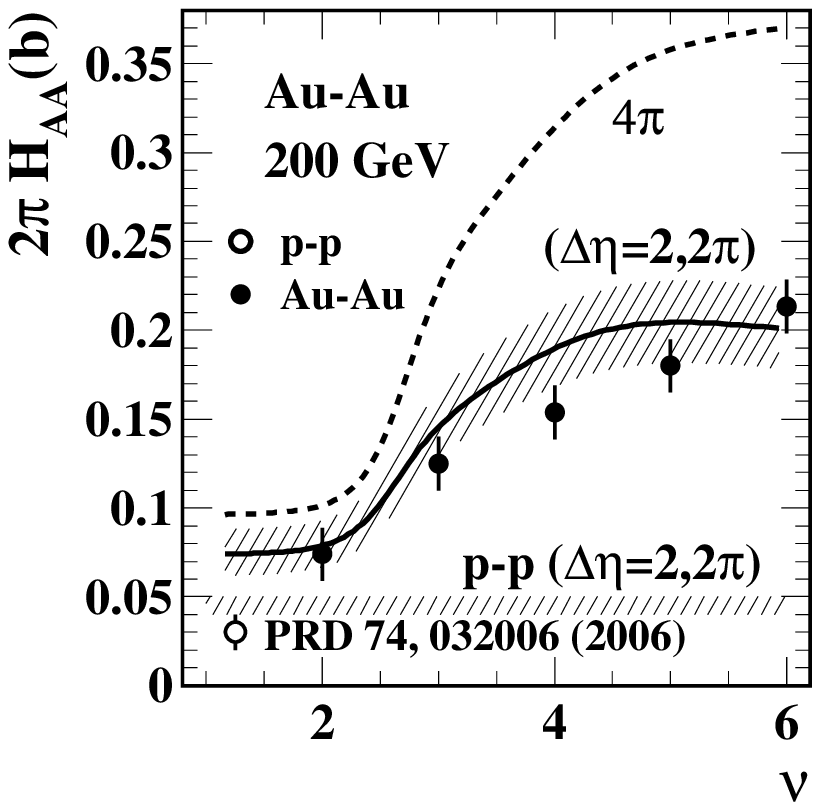} 
   \includegraphics[width=.24\textwidth,height=0.25\textwidth]{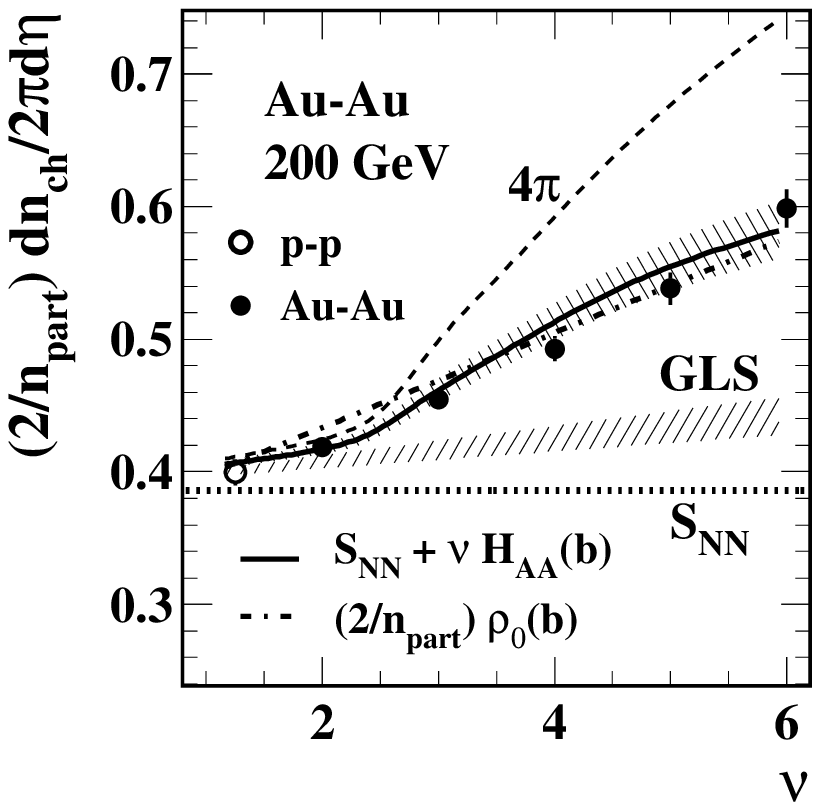} 
\caption{\label{trainor_tom.fig2}
Left panels: 2D angular correlations from 200 GeV \auau collisions  
for (resp.)
 more-peripheral
and more-central collisions~\cite{daugherity}.
Third: Hard-component yield inferred from jet correlations (solid curve) compared to spectrum integrals (points).
Fourth: Two-component total yield inferred from jet correlations (solid curve) compared to spectrum integrals (points).
} 
 \end{figure*}


We want to determine the relation between minimum-bias jets (minijets) and the spectrum hard component (fragmentation). We integrate Eq.~(\ref{fold}) over $y_t$ to obtain the hard-component yield
\bea \label{nchjj}
 2\pi  H_{AA}(b)   &\approx&    \left\{ \frac{\sigma_{dijet}\,\epsilon(\Delta \eta)/2}{ \sigma_{_{\tiny NSD}}\, \Delta \eta_{4\pi}} \right\} \left\{\frac{1}{\sigma_{dijet}}  \int_0^\infty   dy_{max}\, n_{ch,j}(y_{max},b) \frac{d\sigma_{dijet}}{dy_{max}} \right\}
\eea
or $2\pi  H_{AA}(b) =   f(b)\,n_{ch,j}(b)$, where the first factor is the pQCD jet frequency per NSD \pp collision~\cite{ppprd} and the second factor is the minijet mean fragment multiplicity~\cite{jetspec}. Jet angular correlations (same-side jet peak integrals in Fig.~\ref{trainor_tom.fig2} -- first two panels) can be represented by mean pair ratio $j^2(b)$ averaged over angular acceptance $(\Delta \eta,2\pi)$~\cite{jetspec,daugherity}. In terms of  pQCD mean jet number $n_j(b) = n_{bin}(b) \Delta \eta f(b)$ and total multiplicity $n_{ch}(b)$ per \aa collision in $\Delta \eta$ we can write $ n_{ch,j}(b) = n_{ch}(b)\, \sqrt{j^2(b) / n_j(b)}$, which reveals the centrality dependence of the jet fragment multiplicity in \aa collisions.

 Figure~\ref{trainor_tom.fig2} (third panel) shows spectrum hard component $H_{AA}(b)$ (solid curve) inferred from jet angular correlations via Eq.~(\ref{nchjj}). The open point is a spectrum estimate from Ref.~\cite{ppprd} (with $\Delta \eta = 1$). The solid points are derived from the ``total hadrons'' spectrum data in Fig.\ 15 (left panel) of Ref.~\cite{hardspec}.
Multiplying Eq.~(\ref{nchjj}) through by $\nu / 2\pi$ gives
\bea  \label{fraction}
\nu H_{AA}(b) &=& \frac{2}{n_{part}}n_j(b)\frac{n_{ch,j}(b)}{2\pi \Delta \eta}
=  \frac{2}{n_{part}} \rho_0(b)  \sqrt{n_j(b)\, j^2(b)}.
\eea
$\nu H_{AA}(b)$ is the hard component in the two-component spectrum model of Eq.~(\ref{aa2comp}). Figure~\ref{trainor_tom.fig2} (fourth panel) shows the two-component particle yield $S_{NN} + \nu H_{AA}(b)$ predicted by {\em measured} jet angular correlations (bold solid curve). Soft component $s_{NN}$ is by hypothesis fixed at $\sim 0.4$ [2D density on $(\eta,\phi)$] for all \aa centralities.  The solid points are the ``total hadrons'' data in Fig.~15 (left panel) of Ref.~\cite{hardspec} divided by $2\pi$. Minimum-bias angular correlations thus demonstrate that 1/3 of the final state in central \auau collision is contained in resolved jets.

\section{Spectrum structure and paradigm tests}

\paragraph{Jet quenching and $\bf R_{AA}$}

 \begin{figure*}[h] 
  \includegraphics[width=.24\textwidth,height=0.25\textwidth]{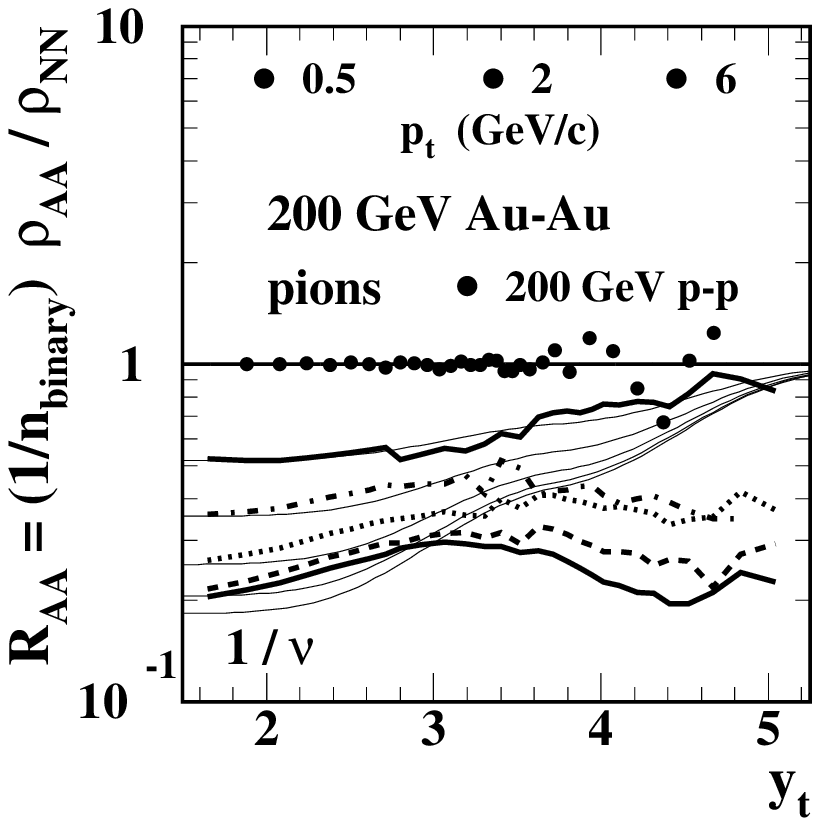}
   \includegraphics[width=.24\textwidth,height=0.25\textwidth]{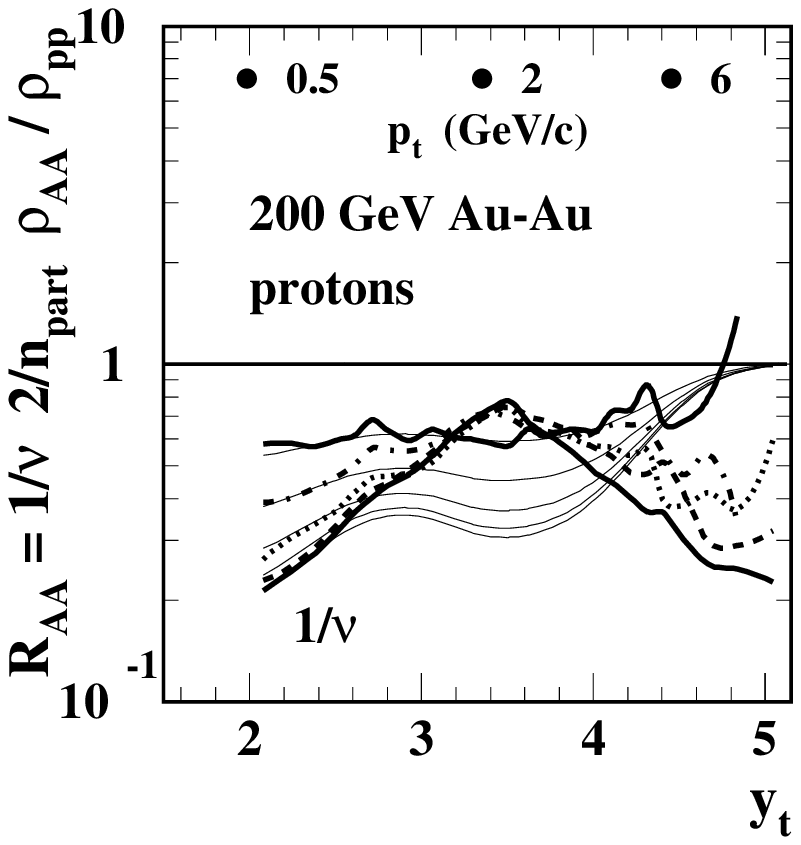} 
  \includegraphics[width=.24\textwidth,height=0.25\textwidth]{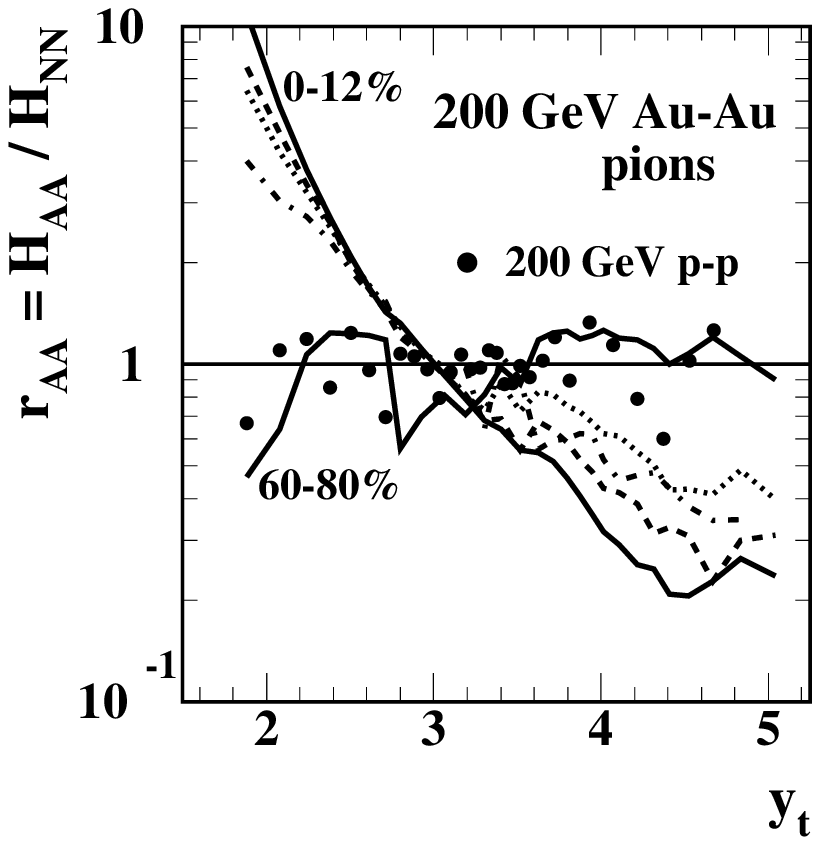}
   \includegraphics[width=.24\textwidth,height=0.25\textwidth]{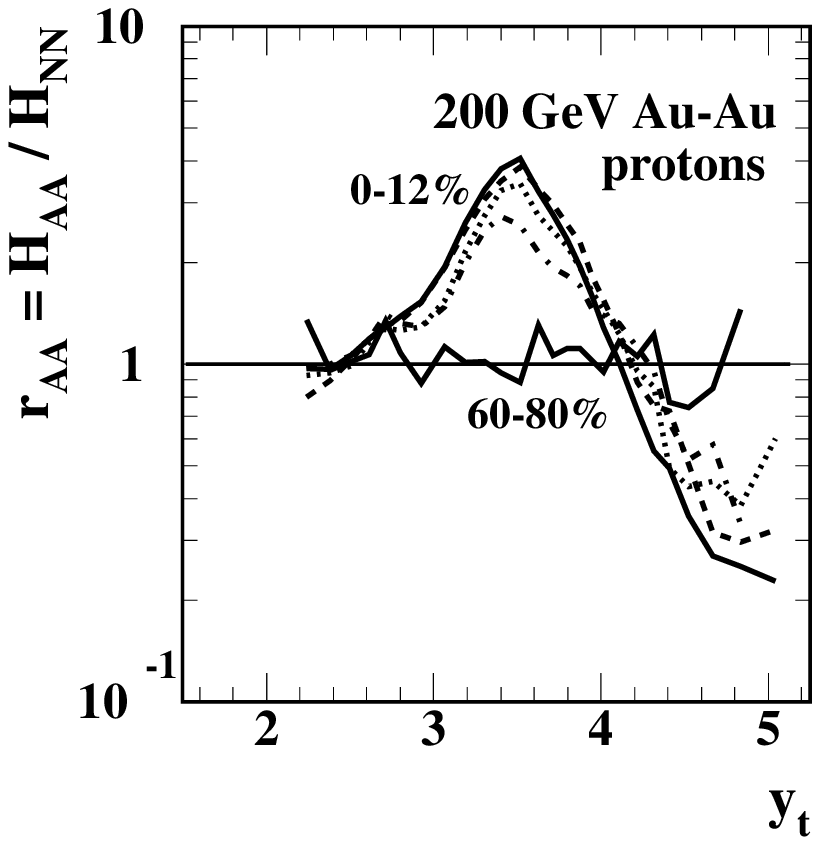} 
\caption{\label{trainor_tom.fig3} 
Left: Conventional spectrum ratios $R_{AA}$ for (resp.) pions and protons from 200 GeV \auau collisions.
Right: Equivalent hard-component ratios $r_{AA}$ for (resp.) pions and protons.
 } 
 \end{figure*}

Spectrum ratio $R_{AA}$ presented as in Fig.~\ref{trainor_tom.fig3} (left panels) suggests strong jet suppression and possible complete absorption of the majority of large-angle scattered partons in more-central \auau collisions at RHIC~\cite{starraa}. However, $R_{AA}$ presents a misleading picture because the ratio includes spectrum soft component $S_{NN}$.  In Fig.~\ref{trainor_tom.fig3} (right panels) alternative {\em hard-component} ratio $r_{AA}$ reveals the true evolution of parton fragmentation with \aa centrality, demonstrating that suppression at larger $p_t$ is compensated by much larger enhancement at smaller $p_t$. The number of {\em jet-correlated} hadrons (hard-component multiplicity) in more-central \aa collisions increases dramatically compared to \pp collisions.

\paragraph{Radial flow}

Radial flow is inferred by so-called blast-wave fits to $p_t$ spectra~\cite{blast}. Such fits generally span a limited $p_t$ interval below 2 GeV/c attributed to ``soft physics'' described by hydro models. But {\em most} of  the parton fragments from jets fall below 2 GeV/c. It can be demonstrated that the spectrum evolution attributed to radial flow is quantitatively predicted by pQCD~\cite{hardspec,fragevo}. Inferred blast-wave parameters actually follow minijet systematics~\cite{nohydro}.

\paragraph{The baryon/meson ratio}

The baryon/meson (B/M) ratio in more-central \aa collisions is said to be anomalous in a pQCD context, suggesting that constituent-quark coalescence from a thermalized partonic medium may provide the dominant particle-production mechanism at intermediate $p_t$~\cite{consquark}. However, examination of corresponding spectrum hard components suggests that parton fragmentation remains the fundamental hadronization process.

 \begin{figure*}[h] 
  \includegraphics[width=.24\textwidth,height=0.25\textwidth]{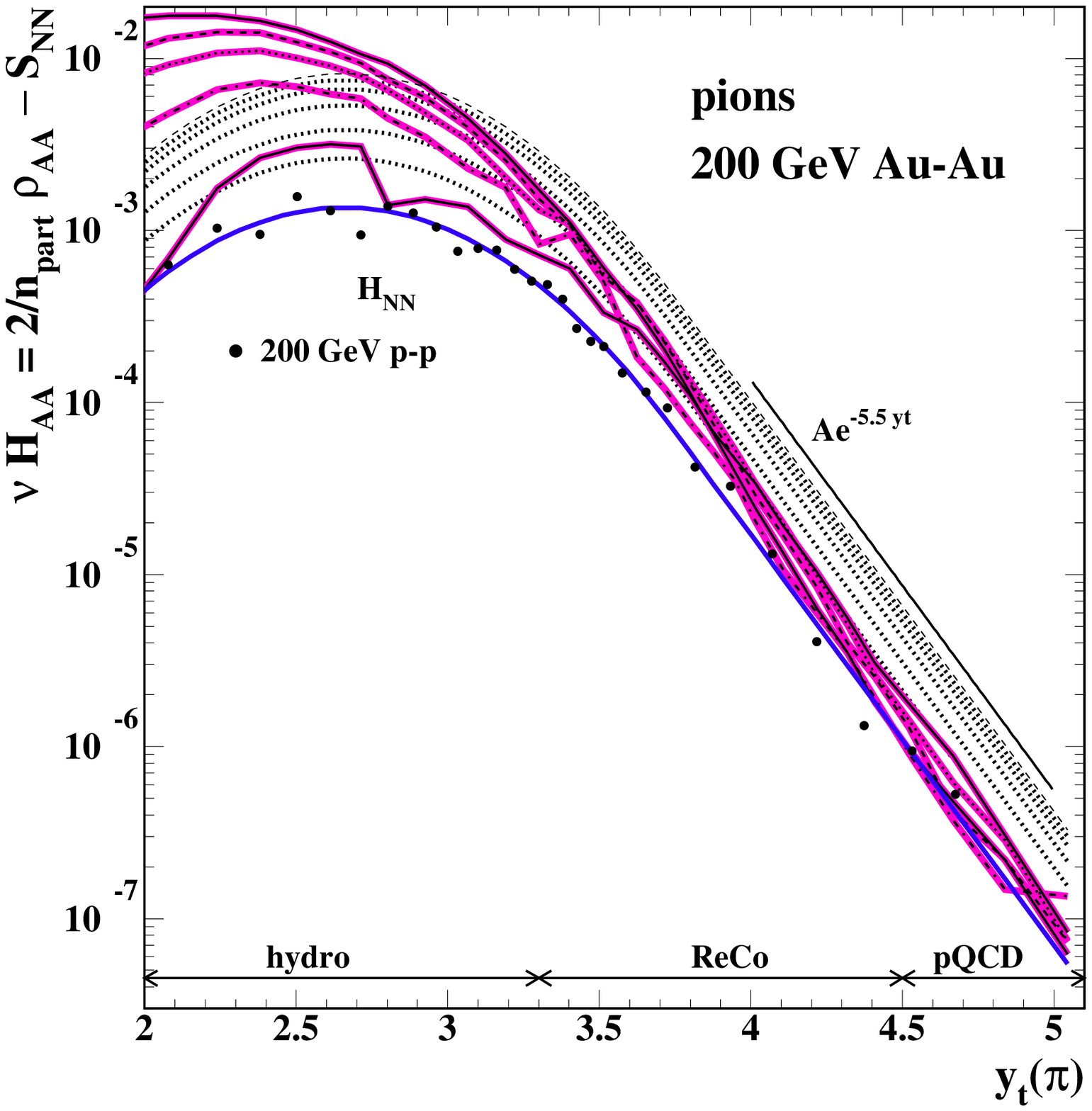}
   \includegraphics[width=.24\textwidth,height=0.25\textwidth]{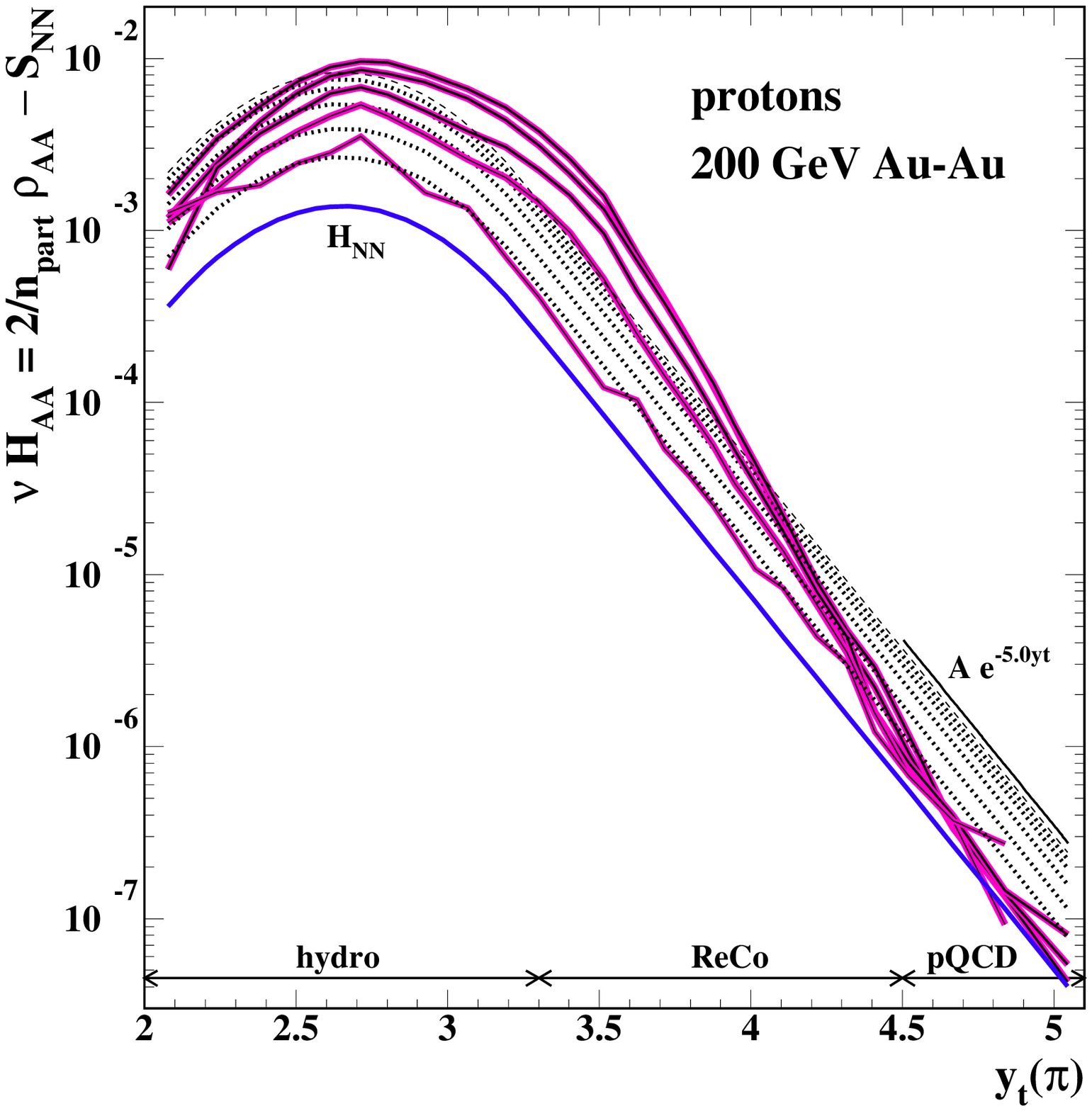}
  \includegraphics[width=.24\textwidth,height=0.25\textwidth]{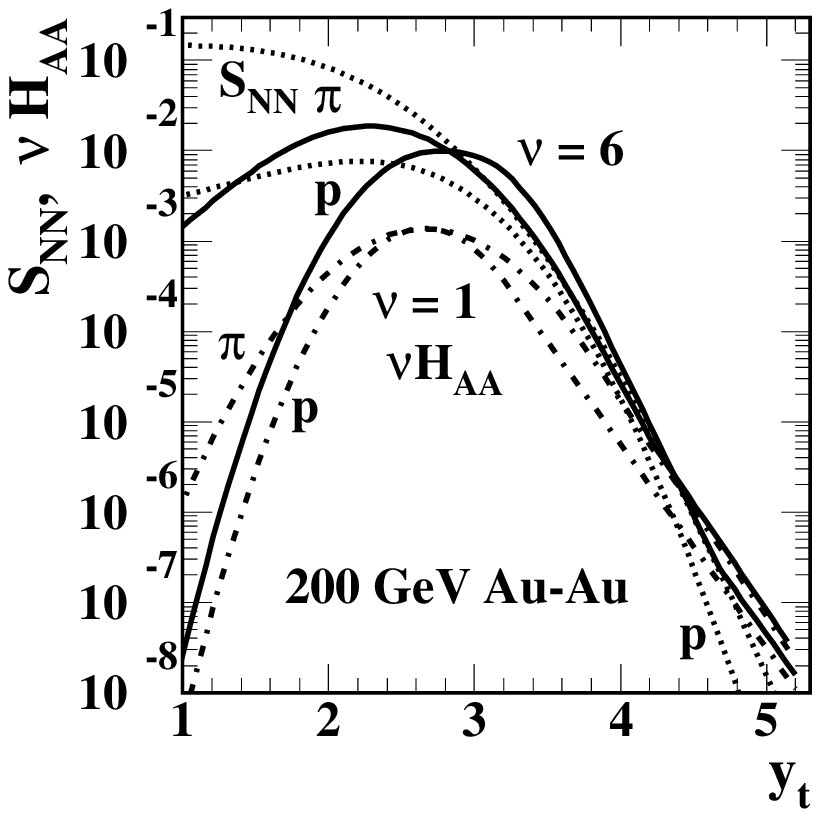} 
   \includegraphics[width=.24\textwidth,height=0.25\textwidth]{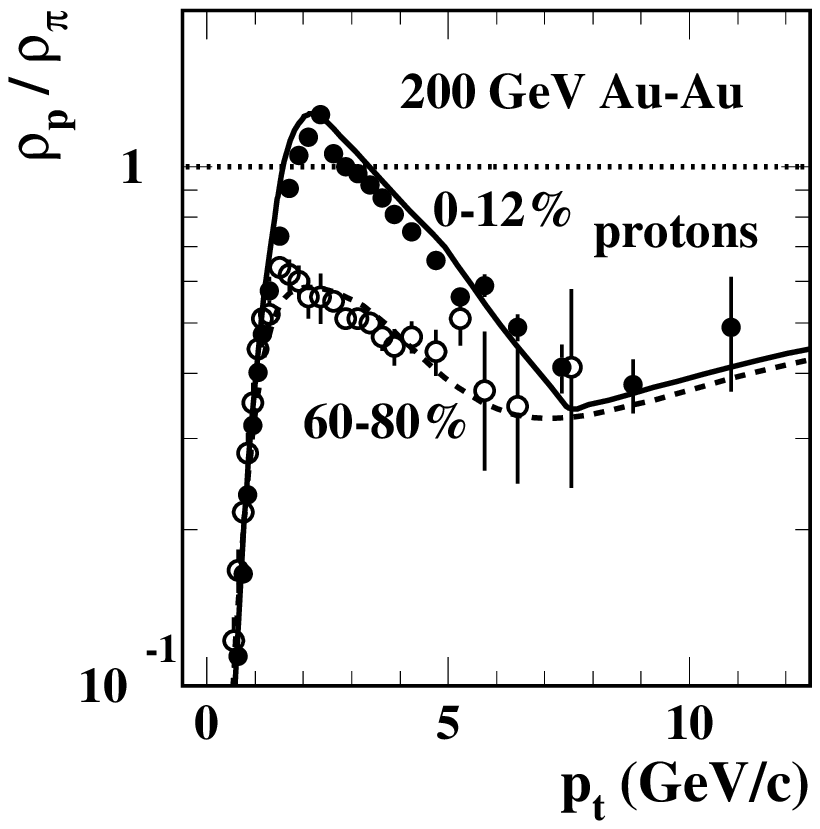} 
\caption{\label{trainor_tom.fig4} 
Left: Spectrum hard components for five centralities of 200 GeV \auau collisions for (resp.) pions and protons.
Third: Soft and hard components for pions and protons from peripheral ($\nu=1$) and central ($\nu=6$) \auau collisions.
Fourth: Proton/pion spectrum ratios.
 } 
 \end{figure*}

Figure~\ref{trainor_tom.fig4} (left panels) shows spectrum hard components for identified pions and protons. The modes for both species in peripheral collisions are located at $p_t = 1$ GeV/c reflecting a common underlying parton spectrum. For more-central collisions the pion mode moves down to $\sim 0.5$ GeV/c {\em but the proton mode remains fixed} on $p_t$~\cite{hardspec}. Both changes can be seen as FF modification as proposed in Ref.~\cite{bw}, but differently controlled in each case by the hadron fragment mass. Figure~\ref{trainor_tom.fig4} (third panel) shows two-component parametrizations of pion and proton spectra accurate to about 10\%. The parameterizations lead to proton/pion spectrum ratios (solid and dashed curves, fourth panel) which accurately describe the B/M ``anomaly'' data, details of which then correspond exactly to parton fragment distributions in the left panels. The structure in the fourth panel correspond exactly to that in Fig.~\ref{trainor_tom.fig3} (right panels).

\section{Correlation structure and paradigm tests}

\paragraph{Dihadron azimuth correlations}

Dihadron azimuth correlations are intended to probe in-medium jet modifications in \aa collisions. A critical issue is subtraction of the combinatoric background modulated by an azimuth quadrupole measured by $v_2$ and conventionally interpreted as ``elliptic flow.'' The absolute background offset is estimated by the ZYAM method, and $v_2$ is obtained from published data derived from {\em nongraphical numerical methods}.
ZYAM subtraction leads to two principal conclusions: (i) most partons are thermalized in an opaque medium and (ii) the development of Mach cones by parton passage through the medium leads to distortion of the away-side azimuth peak (double humps)~\cite{starzyam}. Close examination of the ZYAM subtraction method reveals that the ``zero-yield-at-minimum'' background estimate is not valid for the overlapping same-side and away-side jet peaks encountered in more-central \aa collisions. And published $v_2$ data are typically overestimated due to strong jet (nonflow) contributions~\cite{tzyam,nohydro}. In effect, jet components are subtracted from other jet structure, and jet peak amplitudes are underestimated by an artifically high background offset estimate.


\paragraph{Azimuth quadrupole}

The azimuth quadrupole conventionally interpreted as elliptic flow is measured by quantity $v_2$ which can be defined in terms of nongraphical numerical methods~\cite{newflow}. Alternatively, $v_2\{2D\}$ can be obtained by fitting 2D angular correlations, and novel systematic trends then emerge~\cite{davidhq}. Substantial differences between the two methods can be attributed to strong jet contributions to the former~\cite{davidhq2,davidaustin}. The differences have implications for ZYAM subtraction and Mach-cone inferences~\cite{nohydro}.
The resulting energy and centrality trends of $v_2\{2D\}$ suggest that interpretation of the azimuth quadrupole as a hydro phenomenon is questionable. 

 \begin{figure*}[h] 
 \includegraphics[width=.24\textwidth,height=0.25\textwidth]{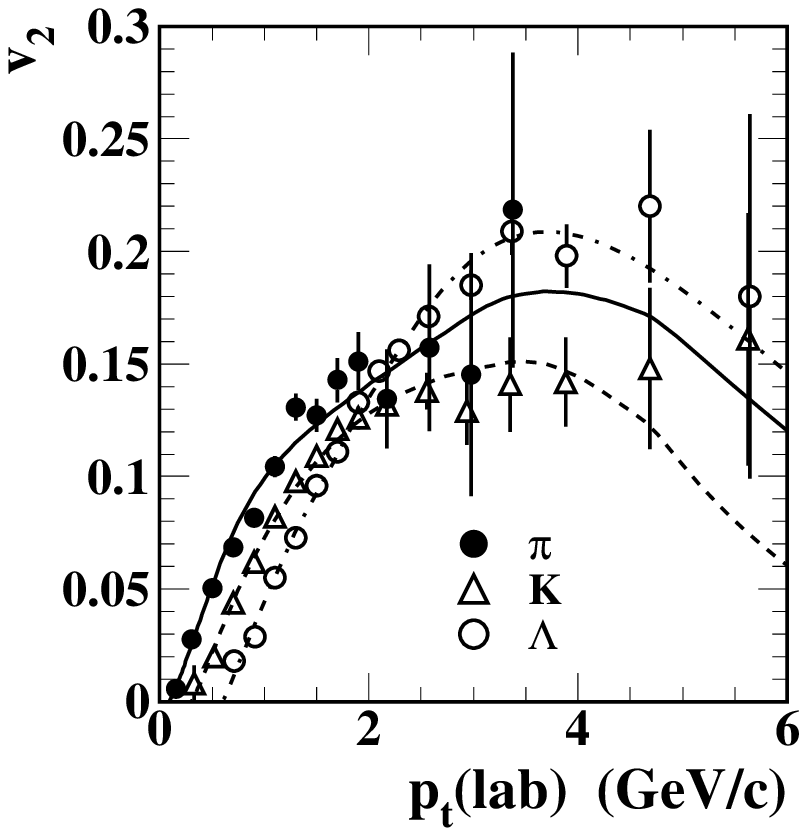}
    \includegraphics[width=.24\textwidth,height=0.25\textwidth]{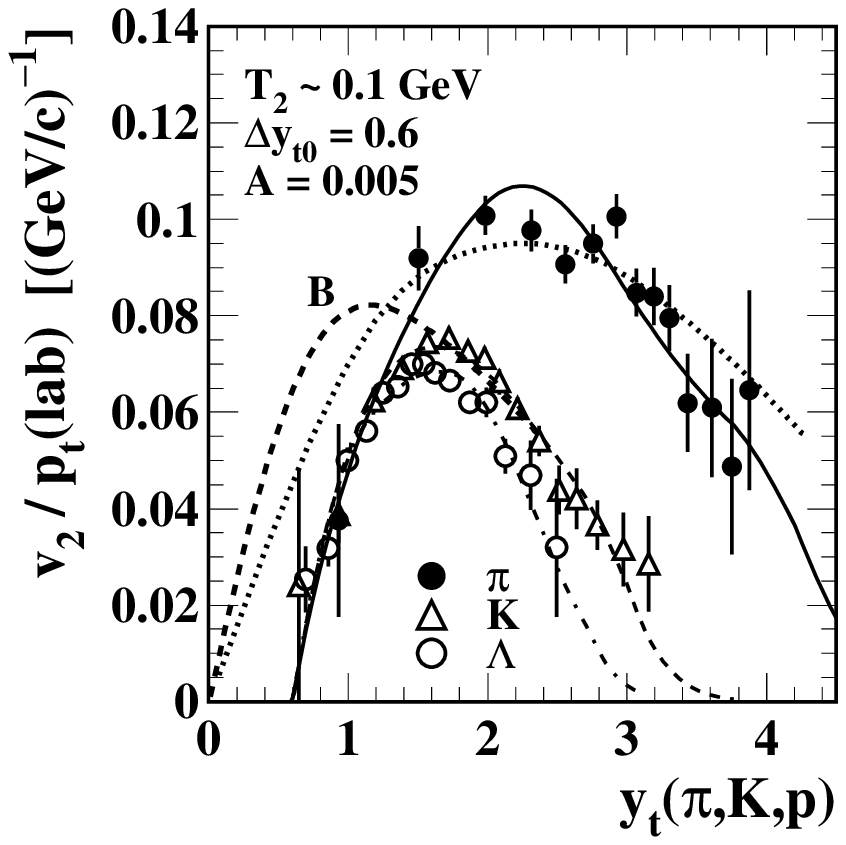} 
  \includegraphics[width=.24\textwidth,height=0.25\textwidth]{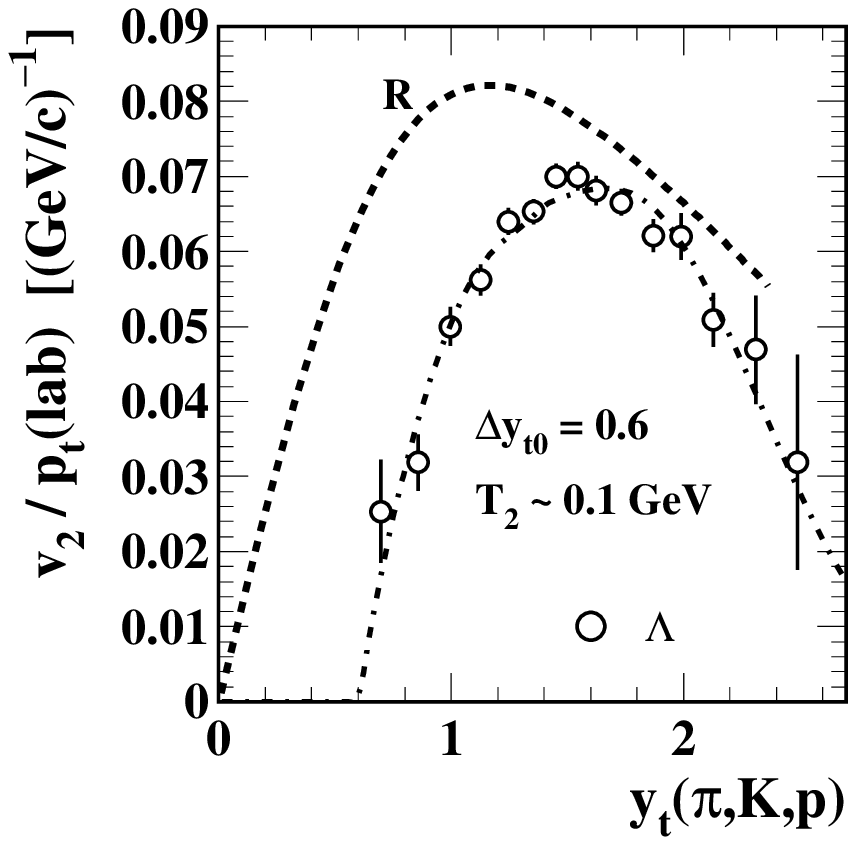}
  \includegraphics[width=.24\textwidth,height=0.245\textwidth]{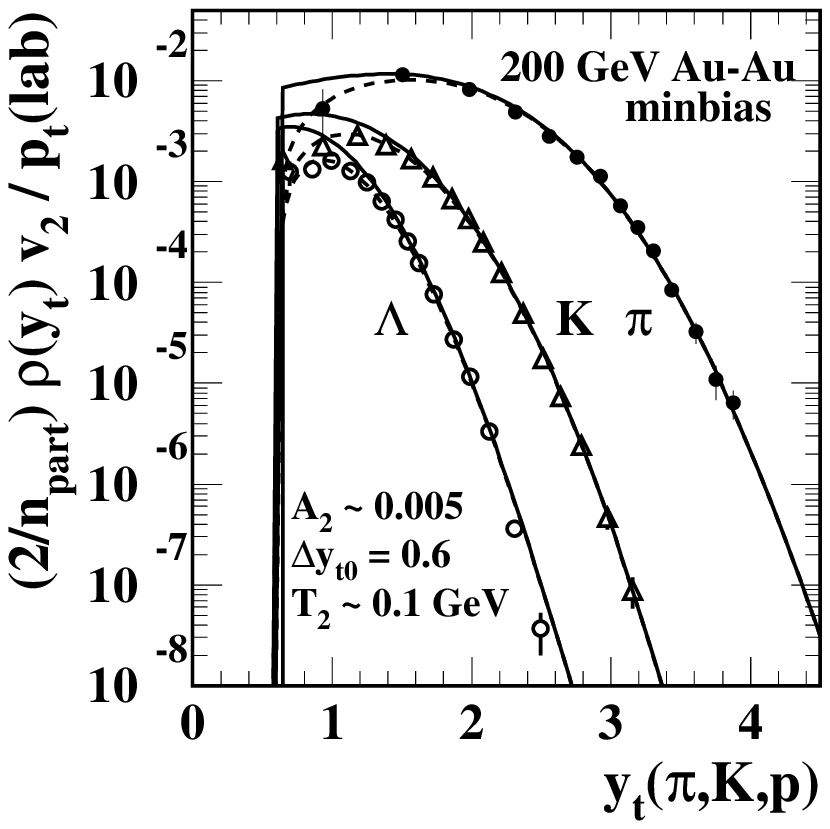} 
\caption{\label{trainor_tom.fig5} 
First: Published $v_2(p_t)$ data for three hadron species.
Second: The same data in a different plotting format.
Third: The same for $\Lambda$s only.
Fourth: Reconstructed quadrupole spectra for the three hadron species, illustrating a common source boost and emission parameters.
 } 
 \end{figure*}

In Fig.~\ref{trainor_tom.fig5} the same published minimum-bias $v_2(p_t)$ data for three hadron species are plotted in several formats. Plotted as $v_2(p_t) / p_t$ on transverse rapidity $y_t = \ln[(p_t + m_t)/m_h]$ with hadron mass $m_h$ (middle panels) the $v_2$ data reveal an apparent common source boost $\Delta y_{t0} \sim 0.6$ and large deviations from a viscous-hydro theory based on Hubble expansion (R)~\cite{quadspec}. When further multiplied by the single-particle spectrum $\rho_0(y_t)$ the $v_2(p_t)$ data (fourth panel) reveal a common {\em quadrupole spectrum} in the form of a L\'evy distribution on $m_t$, much ``colder'' than the single-particle spectrum for most hadrons~\cite{davidhq2,davidaustin}. Those results strongly suggest that ``elliptic flow'' in the form of the azimuth quadrupole is associated with a small fraction of the hadronic final state and results from QCD field-field interactions of low-$x$ gluon condensates~\cite{gluequad}.

\section{Summary}

Hydro analysis of RHIC data interprets particle production below 2 GeV/c in terms of flow phenomena. The role of parton scattering and fragmentation is minimized. But model-independent analysis of spectrum and correlation structure reveals new fragmentation features quantitatively described by pQCD.
$p_t$ spectrum hard components are manifestations of minimum-bias parton fragmentation quantitatively matched to minimum-bias jet angular correlations (minijets). Calculated pQCD fragment distributions accurately describe measured hard components. 

The reference for all fragmentation in nuclear collisions is the FD derived from {\em measured} in-vacuum $e^+$-$e^-$ FFs and the parton (dijet) spectrum for \pp collisions. 
Relative to the reference the spectrum hard component for \pp and peripheral \auau collisions is found to be {\em strongly suppressed} for smaller momenta. At a specific point on \auau centrality the hard component transitions to strong enhancement at smaller fragment momentum and suppression at larger momentum.

Minimum-bias jet (minijet) correlations have been converted to absolute fragment yields  which are found to comprise approximately one third of the final state in central 200 GeV \auau collisions, implying that almost all large-angle scattered partons down to 3 GeV parton energy survive as jet manifestations in the final state, albeit with some modification.
pQCD calculations should be applied to all aspects of spectrum and correlation data to discover what is truly novel in RHIC collisions. pQCD describes almost all RHIC collision evolution---hydro interpretations are questionable.

I greatly appreciate the hospitality of the ISMD organizing committee. This work was supported in part by the Office of Science of the U.S. DOE under grant DE-FG03-97ER41020.



\begin{footnotesize}

\end{footnotesize}


\end{document}